\documentclass[aps,prd,amsmath,amsfonts,amssymb,eqsecnum,nofootinbib,floatfix,secnumarabic,preprintnumbers,superscriptaddress,nobalancelastpage,onecolumn,notitlepage,biblatex]{revtex4-2}

\usepackage{epsfig}
\usepackage{graphicx}
 \usepackage{placeins}
\usepackage{mathrsfs}
\usepackage{amssymb}
\usepackage{float}
\usepackage{verbatim}
\usepackage{amsmath}
\usepackage[toc,page]{appendix}
\usepackage[colorlinks=true,linkcolor=blue,linktocpage=true]{hyperref}
\usepackage{orcidlink}

\expandafter\ifx\csname package@font\endcsname\relax\else
 \expandafter\expandafter
 \expandafter\usepackage
 \expandafter\expandafter
 \expandafter{\csname package@font\endcsname}%
\fi

\begin{document}

\title{\boldmath Primordial black hole evaporation in a thermal bath and gravitational waves}
\author{Arnab Chaudhuri\,\orcidlink{0000-0002-6784-1360}}
\email{arnab.chaudhuri@vit.ac.in}
\affiliation{Department of Physics, School of Advanced Sciences,
Vellore Institute of Technology, Vellore, Tamil Nadu 632014, India.}
\author{Kousik Loho\,\orcidlink{0000-0001-6330-9505}}
\email{kousikloho@hri.res.in}
\affiliation{Regional Centre for Accelerator based Particle Physics, Harish-Chandra Research Institute, A CI of Homi Bhabha National Institute, Chhatnag Road, Jhunsi, Prayagraj 211019, India}

\begin{abstract}
Primordial black holes (PBHs) formed in the early Universe evaporate via Hawking radiation and constitute a generic source of stochastic gravitational waves. Existing studies of gravitational wave production from evaporating PBHs typically assume vacuum evaporation, neglecting the fact that PBHs in the early Universe are embedded in a hot thermal plasma. In this work, we investigate gravitational wave production from primordial black holes whose evaporation is thermally influenced by their surrounding environment. We adopt a thermal evaporation framework in which interactions with the ambient plasma modify the effective decay rate of the black hole, leading to enhanced mass loss at early times and a redistribution of the evaporation history compared to the standard non-thermal vacuum case. Since graviton emission is intrinsically tied to the evaporation history of PBHs, these thermal effects play a crucial role in determining the timing and spectral properties of the resulting stochastic gravitational wave background. Our results provide a consistent framework for incorporating thermal effects into gravitational wave production from evaporating primordial black holes and set the stage for a detailed analysis of their observational signatures.
\end{abstract}
\begin{flushright}
\small{HRI-RECAPP-2026-03}
\end{flushright}
\maketitle
\tableofcontents
\flushbottom
\newpage
\section{Introduction}
\label{sec:intro}

Primordial black holes (PBHs) are a well-motivated consequence of early-Universe dynamics and provide a unique window into physics at energy scales far beyond the reach of terrestrial experiments \cite{Zeldovich:1967lct,Hawking:1971ei,Carr:1974nx,Carr:1975qj}. Formed from the gravitational collapse of large primordial density fluctuations, bubble collisions, or topological defects, PBHs can span an enormous range of masses, from sub-gram scales to astrophysical masses \cite{Hawking:1971ei,Carr:2010wk,Carr:2020gox}. Their formation mechanisms are tightly connected to inflationary physics, reheating dynamics, and possible departures from standard cosmology \cite{Garcia-Bellido:1996mdl,Garcia-Bellido:2017mdw,Pi:2017gih,Byrnes:2018txb,Young:2019yug}.

An important property of PBHs is their evaporation via Hawking radiation \cite{Hawking:1974rv,Hawking:1975vcx}, leading to the emission of all particle species kinematically accessible at the black hole temperature. Depending on their initial mass, PBHs may evaporate during radiation domination, matter domination, or even at late cosmological times \cite{MacGibbon:1991tj,MacGibbon:1990zk}. The associated injection of energy into the primordial plasma can affect a variety of cosmological observables related to big bang nucleosynthesis (BBN), the cosmic microwave background (CMB), and the diffuse gamma-ray background. These effects have been used to derive stringent constraints on the abundance of PBHs across a wide range of masses, under assumptions about the evaporation history and thermalization of the emitted radiation \cite{Carr:2010wk,Kohri:2014lza,Kokubu:2018fxy,Carr:2020gox}.

Beyond their impact on standard cosmological observables, PBHs have attracted renewed interest as sources of new particles and radiation. Hawking evaporation provides a natural production mechanism for dark matter, dark radiation, and feebly interacting particles, with important implications for physics beyond the Standard Model \cite{Fujita:2014hha,Masina:2020xhk,Laha:2020ivk,Hooper:2020evu}. In particular, PBHs offer a unique laboratory for probing gravitational degrees of freedom, since graviton emission is a universal and unavoidable feature of Hawking radiation \cite{Page:1976df,Page:1977um}.

The emission of gravitons from evaporating PBHs leads to a stochastic gravitational wave (GW) background extending over a wide range of frequencies \cite{Anantua:2008am,Barrau:2003xp}. Unlike astrophysical GW sources, this signal originates from the early Universe and carries direct information about PBH mass evolution, evaporation history, and cosmological expansion. Several works have computed the expected GW spectrum from PBH evaporation, exploring its detectability and its role as a complementary probe of PBH populations \cite{Anantua:2008am,Dolgov:2011cq,Dong:2015yjs,Inomata:2020lmk,Perez-Gonzalez:2020vnz}.

However, existing calculations of GW production from PBH evaporation rely almost exclusively on the assumption that black holes evaporate in zero temperature vacuum. Under this assumption, the mass loss rate is determined solely by the Hawking temperature and greybody factors, and the surrounding environment plays no dynamical role \cite{Page:1976df,MacGibbon:1991tj}. While this approximation is valid at late times or for sufficiently hot black holes, it is not generically applicable in the early Universe. In realistic cosmological settings, PBHs are embedded in a hot and dense thermal bath composed of Standard Model particles. When the ambient plasma temperature is comparable to or exceeds the Hawking temperature, the interaction between the black hole and the surrounding thermal environment modifies the effective evaporation dynamics. In this regime, the black hole can no longer be treated as an isolated system radiating into non-thermal vacuum, and its decay must be described within a thermal framework that consistently accounts for the presence of the surrounding plasma \cite{Kalita:2025foa,Chatterjee:2025wnt}. As a result, the evaporation history of the PBH is redistributed in time relative to the non-thermal case, with a modified decay rate and a nontrivial pattern of early- and late-time mass loss, rather than a conventional instantaneous evaporation event.

The phenomenology of primordial black hole evaporation has been extensively studied under the standard vacuum approximation, including its implications for particle production, dark matter, baryon asymmetry, and other cosmological observables \cite{Hooper:2020evu,Laha:2020ivk,Chaudhuri:2025hen,Chaudhuri:2025qwp,Borah:2024lml,Sanchis:2025awq,Morrison:2018xla, Gondolo:2020uqv, Bernal:2020bjf, Green:1999yh, Khlopov:2004tn, Dai:2009hx, Allahverdi:2017sks, Lennon:2017tqq, Hooper:2019gtx, Chaudhuri:2020wjo, Masina:2020xhk, Baldes:2020nuv, Bernal:2020ili, Bernal:2020kse, Lacki:2010zf, Boucenna:2017ghj, Adamek:2019gns, Carr:2020mqm, Masina:2021zpu, Bernal:2021bbv, Bernal:2021yyb, Samanta:2021mdm, Sandick:2021gew, Cheek:2021cfe, Cheek:2021odj, Barman:2021ost, Borah:2022iym,Chen:2023lnj,Chen:2023tzd,Kim:2023ixo,Gehrman:2023qjn,Calza:2023rjt,Coleppa:2022pnf,Chaudhuri:2023aiv,Fujita:2014hha, Hamada:2016jnq, Morrison:2018xla, Hooper:2020otu, Perez-Gonzalez:2020vnz, Datta:2020bht, JyotiDas:2021shi, Smyth:2021lkn, Barman:2021ost, Bernal:2022pue, Ambrosone:2021lsx,Calabrese:2023key,Calabrese:2023bxz,Gehrman:2022imk,Gehrman:2023esa,Schmitz:2023pfy,Bernal:2021yyb,Bernal:2022swt,Barman:2022pdo,Borah:2024qyo,Chianese:2024nyw,Dolgov:2000ht,Acharya:2020jbv,Papanikolaou:2023cku,Chattopadhyay:2022fwa,Chaudhuri:2025rcs}. Only recently has the role of a thermal environment in modifying PBH evaporation been systematically addressed, with Refs.~\cite{Kalita:2025foa,Chatterjee:2025wnt} providing the consistent framework for black hole evaporation in the presence of a thermal bath. While these works establish how thermal effects reshape the evaporation history and lifetime of PBHs, their implications for gravitational wave production have not been explored. In this work, we study for the first time how thermally corrected PBH evaporation impacts the generation of stochastic gravitational waves.

The presence of a thermal bath is expected to have a direct and nontrivial impact on GW emission from PBHs. Since the graviton production rate depends sensitively on the instantaneous Hawking temperature, any modification to the PBH mass evolution feeds directly into the amplitude, spectral shape, and temporal duration of the resulting GW signal. In particular, thermally regulated evaporation redistributes the mass loss history of the PBH relative to the vacuum case, modifying the timing and duration of efficient graviton emission and updating the characteristic spectrum of the resulting gravitational wave background. A distinctive consequence of this redistribution is the redshifted GW spectrum associated with graviton production at early thermally enhanced mass loss phase, followed by the dominant emission during the final effectively non-thermal evaporation stage. As a result, the stochastic gravitational wave background can exhibit a multi-feature structure with observable differences in the spectral shape compared to the standard non-thermal vacuum evaporation scenario.

In this work, we investigate gravitational wave production from evaporating primordial black holes in the presence of a thermal bath. We describe a consistent framework in which thermal effects modify the effective evaporation rate of primordial black holes, leading to a nontrivial redistribution of their mass loss history and graviton emission profile. Our analysis focuses on PBHs evaporating during radiation domination and systematically quantifies how thermal effects reshape the resulting gravitational wave spectrum. This study provides the necessary bridge between recent treatments of thermally updated PBH evaporation and gravitational wave production, and highlights distinctive signatures that can experimentally discover the thermal bath effects.

The rest of this article is organiszed as follows: we discuss the effect of thermal bath on the rate of PBH mass loss in Sec.~\ref{sec:therm} and then describe the corresponding gravitational wave signal in Sec.~\ref{sec:GW} before finally making the concluding remarks in Sec.~\ref{sec:conc}.

\section{PBH evaporation in the presence of a thermal bath}
\label{sec:therm}

A non-rotating, uncharged PBH of mass $M_{\rm PBH}$ is characterized by the Hawking temperature
\begin{equation}
T_{\rm PBH} = \frac{M_{Pl}^2}{8\pi M_{\rm PBH}} \, ,
\label{eq:T_M}
\end{equation}
which increases monotonically as the black hole loses mass through Hawking radiation \cite{Hawking:1974rv,Hawking:1975vcx}. $M_{Pl}$ is the Planck mass. This inverse relation between temperature and mass implies that lighter PBHs evaporate more rapidly and emit progressively harder radiation as evaporation proceeds.

In the standard vacuum description, the PBH mass decreases due to the emission of all kinematically accessible particle species. The corresponding mass loss rate can be written as
\begin{equation}
\frac{dM_{\rm PBH}}{dt}
=
- \sum_i g_i \, \epsilon_i(M_{\rm PBH}) \,
\frac{M_{\rm Pl}^4}{M_{\rm PBH}^2} \, ,
\label{eq:dMdt_vac}
\end{equation}
where $g_i$ denotes the internal degrees of freedom of species $i$, and $\epsilon_i$ is the evaporation function encoding the spectral distribution of emitted particles. This expression makes explicit the characteristic $M_{\rm PBH}^{-2}$ dependence of the mass loss rate, which ultimately leads to the well known cubic scaling of the PBH lifetime with its initial mass
\begin{equation}
\tau_{\rm PBH}
\approx
\frac{(M_{\rm PBH}^{\rm in})^3}{3 M_{\rm Pl}^4}
\left(
\sum_i g_i \epsilon_i
\right)^{-1} .
\label{eq:lifetime}
\end{equation}
For a particle species $i$ with spin $s_i$ and mass $\mu_i$, the evaporation function is given by \cite{MacGibbon:1990zk,Cheek:2021odj}
\begin{equation}
\epsilon_i
=
\frac{27}{8192\pi^5}
\int_{z_i}^{\infty}
\frac{\psi_{s_i}(x)(x^2 - z_i^2)}{e^x - (-1)^{2s_i}} \, xdx \, ,
\label{eq:epsilon_vac}
\end{equation}
where $z_i = \mu_i/T_{\rm PBH}$ and $\psi_{s_i}(x)$ signifies the ratio of the full greybody factors to the geometric optics limit for a particle species with spin $s_i$ \cite{Ukwatta:2015iba}. This function captures the statistical nature of Hawking emission, including the dependence on the spin and mass of the emitted particle species. For brevity, $\psi_{s_i}(x)$ are neglected and set to unity in the rest of this manuscript, allowing for a cost-effective numerical treatment while preserving the qualitative features of the evaporation process.

In the early Universe, PBHs are immersed in a hot thermal plasma rather than evaporating in isolation. As a result, the emitted Hawking radiation interacts with the surrounding thermal bath, and the evaporation process is modified. The influence of the thermal bath can be incorporated directly at the level of the evaporation function. As the radiation emitted gets thermalised with the bath temperature ($T_b$) the evaporation function can be rewritten following Ref.~\cite{Kalita:2025foa} as
\begin{equation}
\epsilon_i
=
\frac{27}{8192\pi^5}
\int_{z_i}^{\infty}
\frac{(x^2 - z_i^2)\,x}{e^x - (-1)^{2s_i}}
\left[
1 + \frac{e^x + (-1)^{2s_i}}{e^{y x} - (-1)^{2s_i}}
\right]
dx \, ,
\label{eq:epsilon_thermal}
\end{equation}
where $y \equiv T_{\rm PBH}/T_b$. The additional term inside the square brackets encodes the effect of the surrounding thermal environment on the emission process. In the limit $T_b \ll T_{\rm PBH}$, corresponding to late times or sufficiently hot black holes, this expression smoothly reduces to the vacuum result in Eq.~\eqref{eq:epsilon_vac}.

However, using a thermofield dynamics approach it was later realized in Ref.~\cite{Chatterjee:2025wnt} that the modified evaporation function consists of two parts. One part is responsible for the net particle production from PBHs and the other part is the contribution of the bath. It is the first of the two contributions, that is relevant for our study of PBH evaporation. After subtracting the bath contribution the evaporation function can be written as \cite{Chatterjee:2025wnt}

\begin{equation}
\epsilon_i
=
\frac{27}{8192\pi^5}
\int_{z_i}^{\infty}
\frac{(x^2 - z_i^2)\,x}{e^x - (-1)^{2s_i}}
\left[
1 + 2\frac{ (-1)^{2s_i}}{e^{y x} - (-1)^{2s_i}}
\right]
dx \, ,
\label{eq:epsilon_final}
\end{equation}
In what follows, we compare the thermal effects of the bath given by Eq.~\eqref{eq:epsilon_final} to the isolated PBH given by Eq.~\eqref{eq:epsilon_vac}. The inclusion of thermal effects through Eq.~\eqref{eq:epsilon_final} modifies the numerical value of the PBH lifetime relative to the standard vacuum case, while preserving the characteristic cubic scaling with the initial mass. This change in lifetime reflects the altered efficiency of energy loss when Hawking radiation interacts with a thermal environment.

The evolution of the PBHs along with the cosmological background can be classified into two distinct phases. In the early era around the time of the formation of the PBHs, the temperature of the bath could be higher than the Hawking temperature of the PBH ($T_b>T_{PBH}$) where Eq.~\eqref{eq:epsilon_final} is to be used to get the correct rate of evaporation. Over a few efolds the temperature of the universe cools down. However, the temperature of the PBHs rise due to mass loss via Hawking radiation since they are related via Eq.~\eqref{eq:T_M} ultimately leading to a phase where $T_{PBH}>T_b$ and Eq.~\eqref{eq:epsilon_vac} can be used approximately. The bath temperature $T_b$ is identified with the temperature of the radiation-dominated Universe.

\begin{figure}[htp!]
\centering
\includegraphics[scale=0.5]{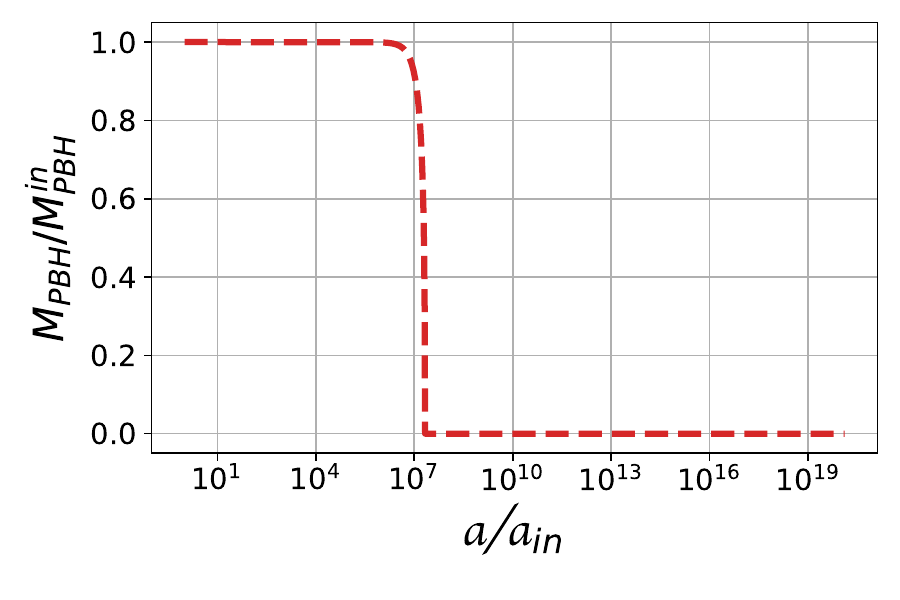}
\includegraphics[scale=0.5]{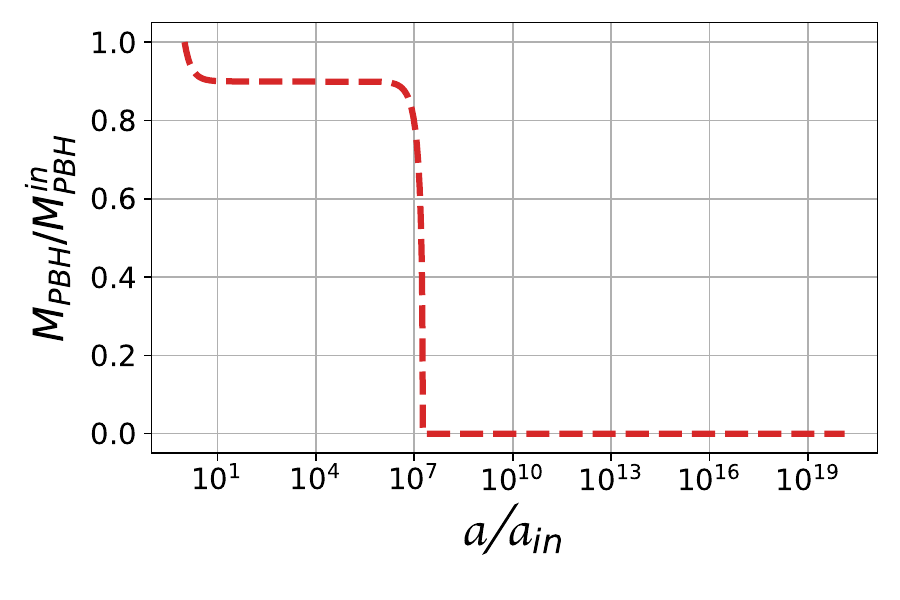}
\caption{\textit{Left:} The evolution of PBH mass with the scale factor without taking the bath temperature into account. The benchmark choices of the relevant parameters are taken to be $M_{in}=10^2$ grams and $\beta=10^{-10}$. \textit{Right:} Same as the left counterpart except that the bath temperature has been taken into account.}
\label{fig:mass}
\end{figure}

The initial PBH mass is related to the formation temperature $T_{\rm in}$ by\footnote{The Refs.~\cite{Kalita:2025foa,Chatterjee:2025wnt} have treated the initial mass of PBH  and their formation temperature as independent parameters. However, we have used Eq.~\eqref{eq:Min_Tin} to estimate $T_{in}$ which leads to a very different mass evaporation profile in Fig.~(\ref{fig:mass}) compared to those references.}
\begin{equation}
M_{\rm PBH}^{\rm in}
=
\frac{4\pi}{3}\,\gamma\,
\frac{\rho_R(T_{\rm in})}{H^3(T_{\rm in})} \, ,
\label{eq:Min_Tin}
\end{equation}
where $\rho_R$ is the radiation energy density, $H$ is the Hubble rate, and $\gamma$ is an ${\cal O}(1)$ collapse efficiency factor. This relation fixes the initial conditions for the subsequent evaporation history in terms of the cosmological background at formation.

The modified evaporation rate implied by Eqs.~\eqref{eq:epsilon_final} and \eqref{eq:lifetime} leads to a nontrivial time evolution of the PBH mass when thermal effects are important. This altered mass evolution changes both the timing and duration of particle emission from PBHs relative to the standard vacuum scenario. The changes are depicted in Fig.~(\ref{fig:mass}), where the evolution of mass has been plotted\footnote{The plots are produced using the publicly available \texttt{FRISBHEE} package \cite{Cheek:2021odj,Cheek:2021cfe,Cheek:2022dbx,Cheek:2022mmy} with some major simplifications.} with the scale factor ($a$). The scale factor at formation is represented by $a_{in}$. The relevant equations are listed in appendix \ref{app:equns}. We have used a monochromatic mass distribution of Schwarzschild PBHs with initial mass $M_{PBH}^{in}=10^2$ grams and have chosen the parameter $\beta=10^{-10}$ in a manner such that the Universe remains radiation dominated throughout the period of PBH evaporation. The temperature of the PBHs at formation is found to be $1\times10^{11}$ GeV using Eq.~\eqref{eq:T_M} and the bath temperature at that instance is found to be $4\times10^{14}$ GeV using Eq.~\eqref{eq:Min_Tin}. Hence, the effect of thermal bath is very prominent at the formation as has been shown on right figure of Fig.~(\ref{fig:mass}) compared to the standard scenario of isolated evaporation shown on the left. The presence of the thermal bath enhances the evaporation around the early phase of the PBHs. Furthermore, a very small reduction in the lifetime of the PBHs has been noticed due to the thermal effects. While, this kind of a double burst feature in the evaporation of PBHs does not change the parameter space associated with the particulate dark matter production significantly, it can still have important consequences for GW emission.

\FloatBarrier

Since graviton emission is an unavoidable component of Hawking radiation, the thermally modified evaporation history plays a central role in determining the resulting gravitational wave signal. In the following section, we use this framework to compute gravitational wave production from evaporating PBHs in presence of a thermal bath.

\section{PBH evaporation and gravitational waves}
\label{sec:GW}
\begin{figure}[htp!]
\centering
\includegraphics[scale=0.5]{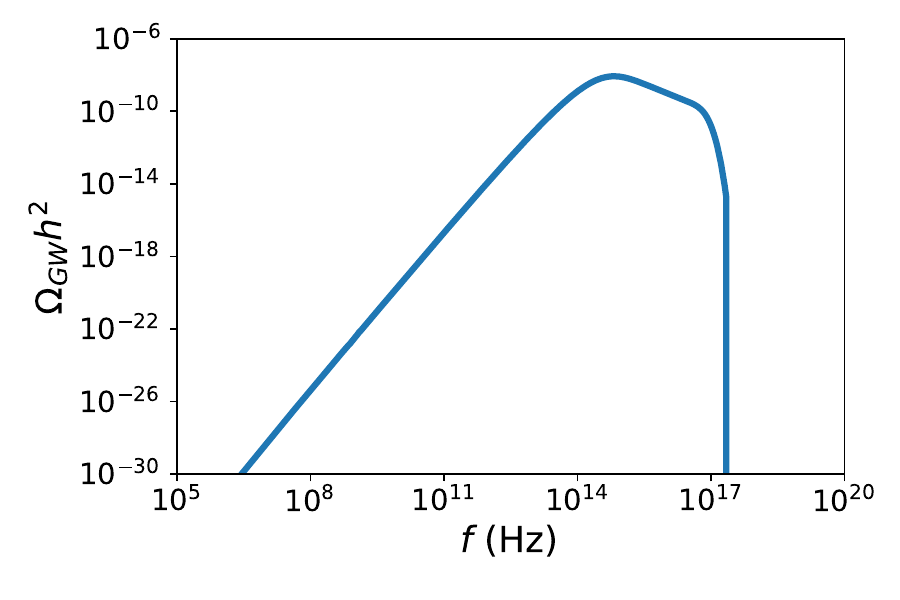}
\includegraphics[scale=0.5]{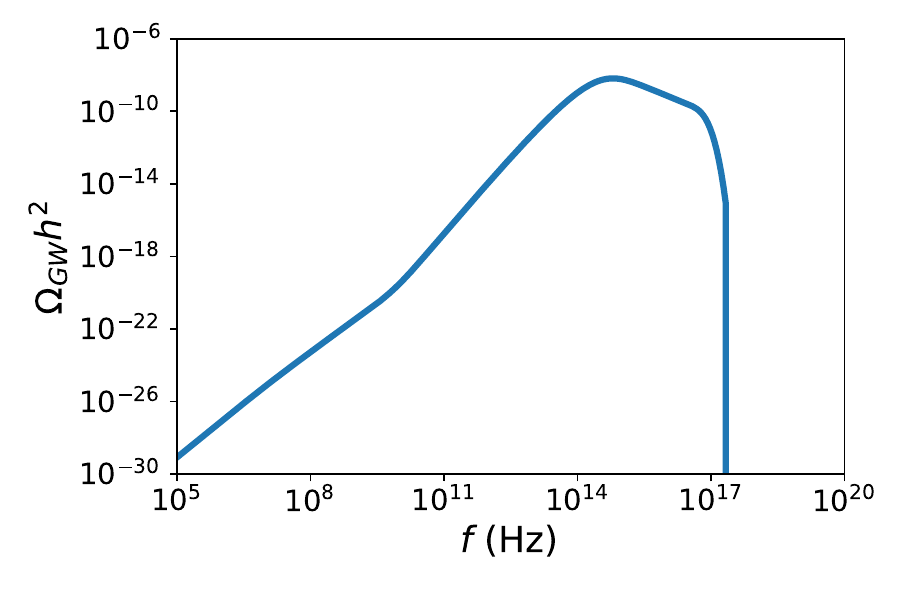}
\caption{\textit{Left:} Gravitational waves produced due to graviton emission during Hawking evaporation of PBH in absence of bath effects. The benchmark choices considered for this figure are $M_{in}=10^2$ grams and $\beta=10^{-10}$. \textit{Right:} Same as the left counterpart apart from the fact that the bath temperature has been taken into account.}
\label{fig:GW}
\end{figure}

Primordial black holes produce all kinds of particles as they evaporate via Hawking radiation and gravitons are of no exception. The emitted gravitons can constitute a source of stochastic gravitational waves \cite{Anantua:2008am,Dolgov:2011cq,Dong:2015yjs,Inomata:2020lmk,Perez-Gonzalez:2020vnz}. The energy density contained in the GW can be obtained by integrating the Hawking radiation spectrum of gravitons over the limetime of the PBHs. The evaporation spectrum roughly follows a Planck distribution upto a greybody factor. The GW relic produced from the evaporation of a monochromatic mass distribution of Schwartschild PBHs can be expressed in terms of frequency ($f$) as the following
\begin{equation}
\Omega_{GW}(f)h^2\approx\frac{g_hf^4n_{PBH}^{in}}{\rho_cM_{pl}^4}\int_{a_{in}}^{a_{ev}}\frac{da}{a^4H(a)}\frac{M_{PBH}^2(a)}{\bigg[\exp{\bigg(\frac{2\pi a_0f}{aT_{PBH}(a)}}\bigg)-1\bigg]},
\label{eq:GW}
\end{equation}
where $g_h$ is the effective degrees of freedom of graviton, $n_{PBH}^{in}$ is the initial number density of PBH, $a_0$ is the scale factor today, $M_{pl}=M_{Pl}/\sqrt{8\pi}$ is the reduced Planck mass, $\rho_c$ is the critical energy density of the Universe, $a_{in}$ is the scale factor at PBH formation and $a_{ev}$ is the scale factor when the PBH completely evaporates. The stochastic GW spectrum, for the benchmark choices same as that of Fig.~(\ref{fig:mass}), is shown on the \emph{left} panel of Fig.~(\ref{fig:GW}). The frequency range of such GWs are at a much higher range compared to the operational range of the present and upcoming conventional GW detectors. However, there have been proposals to probe this regime through inverse Gertsenshtein effect \cite{He:2023xoh}.

Motivated by the double burst feature of mass evaporation profile on the \emph{right} panel of  Fig.~(\ref{fig:mass}), we aim to explore if there exists any observable signatures of such a feature. If a distinct signature can be obtained for such a feature, it could provide an important probe of the effects of thermal bath on PBH evaporation. In this regard, the expression for GW energy density can be updated following Eq.~\eqref{eq:epsilon_final} as 
\begin{equation}
\Omega^{th}_{GW}(f)h^2\approx\frac{g_hf^4n_{PBH}^{in}}{\rho_cM_{pl}^4}\int_{a_{in}}^{a_{ev}}\frac{da}{a^4H(a)}\frac{M_{PBH}^2(a)}{\bigg[\exp{\bigg(\frac{2\pi a_0f}{aT_{PBH}(a)}}\bigg)-1\bigg]}\Bigg[1+\frac{2}{\exp{\bigg(\frac{2\pi a_0f}{aT_{b}(a)}}\bigg)-1}\Bigg],
\label{eq:GW_therm}
\end{equation}
where the effects of thermal bath has also been taken into account while sovling for $M_{PBH}(a)$. With the thermally regulated evaporation function, one would expect another peak in the GW spectrum corresponding to the early burst in the mass evaporation profile. Furthermore, the energy density contained in such GWs will be redshifted for longer than the final burst leading to modifications in the low-frequency behaviour. The GW spectrum from graviton emission of PBH taking thermal bath into account has been shown on the \emph{right} panel of Fig.~(\ref{fig:GW}). It is evident from the figure that the early burst is not strong enough to provide the required resolution for a second peak at low frequencies but only its low frequency tail is visible in this figure. Even though a distinct peak would certainly have been a better probe, one can still derive a great deal of information from the modified shape of the spectrum. If and when better experimental probes of stochastic GWs in such high frequencies become available, one would require to conduct a bayesian analysis to fit the data to provide indications on whether such effects exist in comparison to thermally isolated PBHs without bath effects. The spectral shape of the GW spectrum could be of utmost importance in such an analysis. We have shown in this section that the existence of thermal bath effects can modify the stochastic GW spectrum associated with graviton emission from PBH and thus very high frequency GW experiments can possibly probe this phenomena in future.

\section{Concluding remarks}
\label{sec:conc}

In this work we have investigated the impact of a thermal environment on GW production from evaporating PBHs. Unlike the standard non-thermal vacuum approximation, in which the PBHs evaporate through a runaway process at the end according to Eq.~\eqref{eq:epsilon_vac}, the presence of a thermal bath modifies the effective evaporation function through Eq.~\eqref{eq:epsilon_final}, leading to a redistribution of the PBH mass loss history.

When the bath temperature exceeds the Hawking temperature ($T_b > T_{\rm PBH}$), thermal effects enhance the effective evaporation rate relative to the isolated case. As a consequence, a finite fraction of the PBH mass is radiated away during an early thermally modified phase. As the Universe expands and cools, the condition $T_{\rm PBH} \gtrsim T_b$ is eventually satisfied, after which the evaporation proceeds approximately as in a non-thermal vacuum. This two-stage evolution results in a very slightly reduced lifetime compared to the isolated scenario, as explicitly shown in Fig.~(\ref{fig:mass}).

Since graviton emission is intrinsically tied to the instantaneous Hawking temperature, the modified mass evolution directly reshapes the gravitational wave spectrum. In the thermally updated scenario, graviton production receives contributions from both the early thermally influenced burst and also the usual final burst. However, for the parameter choices considered here, these two stages do not produce clearly separated peaks in the gravitational wave spectrum. Instead, the resulting stochastic GW exhibits a single dominant peak whose shape is subtly distorted relative to the isolated PBH scenario, with a mild enhancement and spectral reshaping toward lower frequencies. As shown in Fig.~(\ref{fig:GW}), thermal corrections therefore imprint a characteristic deformation of $\Omega_{\rm GW}(f)$ rather than generating a fully distinct multi peak structure.

Although the peak frequencies associated with PBHs of mass $M_{\rm in}=10^2\,\mathrm{g}$ lie well beyond the sensitivity range of present and near-future detectors, the spectral distortions induced by thermal regulation are theoretically robust. Should ultra high frequency GWs become experimentally accessible, thermally enhanced PBH evaporation could constitute a distinctive probe of early Universe thermal dynamics.

More broadly, our results demonstrate that incorporating thermal effects is necessary for a self-consistent treatment of PBH evaporation in realistic early Universe environments. While the overall allowed parameter space for subdominant PBHs may remain largely unchanged, the temporal distribution of mass loss and consequently of graviton emission is modified relative to the standard vacuum approximation. For the light PBHs considered here, this redistribution does not generate a qualitatively new multi peak gravitational wave spectrum, but instead induces a measurable deformation of the spectral shape and a shift in the relative power toward lower frequencies. The framework developed in this work therefore establishes how thermal corrections propagate into gravitational wave observables and provides a systematic basis for extending such analyses to other PBH mass ranges and cosmological scenarios.

\acknowledgements
The authors gratefully acknowledge useful discussions with Kazunori Kohri. KL wants to thank Debaprasad Maity for useful discussions.

\appendix
\section{Evolution of energy densities}
\label{app:equns}
The relevant equations that describes the evolution of the energy densities ($\rho_i$) and the bath Temperature ($T_b$) are given by

\begin{equation}
\frac{d\rho_{PBH}}{dt}+3H\rho_{PBH}=\frac{\rho_{PBH}}{M_{PBH}}\frac{dM_{PBH}}{dt},
\end{equation}
\begin{equation}
\frac{d\rho_R}{dt}+4H\rho_R=-\frac{\rho_{PBH}}{M_{PBH}}\frac{dM_{PBH}}{dt},
\end{equation}
\begin{equation}
\frac{dT_b}{dt}=-\frac{T_b}{\Delta}\bigg(H+\frac{1}{M_{PBH}}\frac{dM_{PBH}}{dt}\frac{\rho_{PBH}}{4\rho_R}\frac{g_\star(T_b)}{g_{\star s}(T_b)}\bigg),
\end{equation}
where $\Delta=1+\frac{T_b}{3g_{\star s}(T_b)}\frac{dg_{\star s}(T_b)}{dT_b}$, and $H=\sqrt{\frac{8\pi G(\rho_{PBH}+\rho_R)}{3}}$. The above equations along with Eq.~\eqref{eq:dMdt_vac} can be solved to understand the PBH evaporation.

\bibliography{Reference}
\end{document}